\documentclass[aps,prl,preprintnumbers,twocolumn,groupedaddress,nofootinbib]{revtex4}

\usepackage{latexsym}
\usepackage{amsmath}
\usepackage{amsfonts}

\newcommand{\bq}{\begin{eqnarray}}
\newcommand{\eq}{\end{eqnarray}}
\newcommand{\eps}{\varepsilon}

\newcommand{\arxivdate}{January 18, 2016}

\begin{document}

\preprint{MITP/16-009}
\title{\boldmath{Double copies of fermions as only gravitational interacting matter}}

\author{Leonardo de la Cruz, Alexander Kniss and Stefan Weinzierl}
\affiliation{PRISMA Cluster of Excellence, Institut f{\"u}r Physik, Johannes Gutenberg-Universit\"at Mainz, D-55099 Mainz, Germany}

\date{\arxivdate}

\begin{abstract}
Inspired by the recent progress in the field of scattering amplitudes,  
we discuss hypothetical particles which
can be characterised as the double copies of fermions -- in the same way gravitons can be viewed as 
double copies of gauge bosons.
As the gravitons, these hypothetical particles interact only through gravitational interactions.
We present two equivalent methods for the computation of the relevant scattering amplitudes.
The hypothetical particles can be massive and non-relativistic.
\end{abstract}

\maketitle

\section{Introduction}

It is well-known that there is a close relationship between scattering
amplitudes in pure Yang-Mills theory and gravity.
This relationship can be expressed in several equivalent ways.
One possibility is given by the Kawai-Lewellen-Tye (KLT) relations \cite{Kawai:1985xq},
a second possibility is based on a duality between colour and kinematic numerators \cite{Bern:2010ue}
and coined the expression ``double copy'',
while a third possibility uses the Cachazo-He-Yuan (CHY) representation of amplitudes \cite{Cachazo:2013hca,Cachazo:2013iea}.
In this letter we are interested in extending this relationship on the gauge theory side from 
pure Yang-Mills theory -- consisting of massless gauge bosons (``gluons'') in the adjoint representation of the gauge group --
towards a QCD-like theory by including massless or massive fermions (``quarks'') in the fundamental representation
of the gauge group.
We will study the corresponding extension on the gravity side, consisting of double copies of gluons, quarks and antiquarks.
We discuss two methods for the computation of the gravitational amplitudes and present evidence that they give identical
results.
The first method is based on colour-kinematics duality and incorporates ideas already present in the literature \cite{Bern:2010ue,Chiodaroli:2013upa,Johansson:2014zca,Chiodaroli:2015wal}.
The second method generalises the KLT-relations \cite{Kawai:1985xq,Bern:1999bx,BjerrumBohr:2004wh,BjerrumBohr:2010ta,Feng:2010br,Damgaard:2012fb}
and relates -- similar to the CHY-representation -- gravitational amplitudes, gauge amplitudes 
and three-valent scalar amplitudes.
The generalised KLT relations and the fact that they agree with colour-kinematics duality are new results.
We present a simple formula for the KLT momentum kernel.
Our approach does not rely on supersymmetry nor string theory.

The quarks and the associated double copies may be massless or massive, giving us the opportunity of defining
gravitational amplitudes with massive non-relativistic particles.
These massive particles interact only through gravitational interactions.
We comment on the implications for dark matter.

\section{Notation}

We consider a scattering process with $n$ external particles.
We will assume that $n_q$ of these particles carry flavour (with $ 0 \le n_q \le \lfloor{n/2}\rfloor$)  
and that $n_q$ additional particles carry the corresponding anti-flavour.
We will further assume that the flavour of any pair of flavoured particles is distinct, as is the flavour of any pair
of anti-flavoured particles.
The number of un-flavoured particles is denoted by $n_g$ and clearly we have $n=n_g+2n_q$.
Flavoured particles may be massive, while un-flavoured particles are assumed to be massless.
We consider three types of tree-level amplitudes:
\\
\begin{tabular}{lll}
1. & double-ordered scalar amplitudes & $m_n(p,w,\tilde{w})$, \\
2. & single-ordered gauge amplitudes & $A_n(p,w,\eps)$, \\
3. & un-ordered gravitational amplitudes & $M_n(p,\eps,\tilde{\eps})$. \\
\end{tabular}
\\
The notation for the arguments of the various amplitudes is as follows:
We denote the $n$-tuple of external momenta by $p=(p_1,...,p_n)$
and an external order (being a permutation of $(1,...,n)$) by $w$.
The single-ordered gauge amplitudes $A_n$ are the well-known
tree-level primitive QCD amplitudes.
These depend also on external polarisations, i.e.
polarisation vectors $\eps_j$ for external gluons,
spinors $\bar{u}_j$ for out-going fermions and spinors $v_j$ for out-going anti-fermions.
Unless stated otherwise we will assume all particles to be out-going.
We denote the $n$-tuple of external polarisations by $\eps$.
The double-ordered scalar amplitudes $m_n$ depend on two orderings $w$ and $\tilde{w}$, the
un-ordered gravitational amplitudes $M_n$ depend on two $n$-tuples of polarisations $\eps$ and $\tilde{\eps}$.
The polarisation of a double copy of a gluon is described 
by a product of spin-$1$ polarisation vectors.
The equal spin cases
$\eps_\mu^\pm \eps_\nu^\pm$
correspond to the two polarisations of a single graviton, while
the opposite spin cases
$\eps_\mu^\pm \eps_\nu^\mp$
corresponds to linear combinations of a dilaton and an antisymmetric tensor.
If an external state in $A_n$ corresponds to a quark, 
then the polarisation of the corresponding external state in $M_n$ will be described by
\bq
 \bar{u}_\alpha^{\lambda} \bar{u}_\beta^{\tilde{\lambda}}.
\eq
This motivates the name ``double copy of a fermion'' in the title.
Again, the spins may be equal
$\bar{u}_\alpha^\pm \bar{u}_\beta^\pm$, or opposite $\bar{u}_\alpha^\pm \bar{u}_\beta^\mp$,
leading to $2+2=4$ polarisation states.
In a similar way, if an external state in $A_n$ corresponds to an anti-quark, 
then the polarisation of the corresponding external state in $M_n$ will be described by
$v_\alpha^{\lambda} v_\beta^{\tilde{\lambda}}$,
which may be dubbed ``double copy of an anti-fermion''.

We review some properties of the primitive QCD amplitudes $A_n$.
For fixed $n$ and a given flavour assignment there are relations among the amplitudes $A_n$ with different external orders $w$,
consisting of cyclic invariance, 
the Kleiss-Kuijf relations \cite{Kleiss:1988ne},
the ``no-crossed-flavour-lines''-relations and 
the Bern-Carrasco-Johansson (BCJ) relations \cite{Bern:2008qj,BjerrumBohr:2009rd,Stieberger:2009hq,Feng:2010my,Naculich:2014naa,Johansson:2015oia,delaCruz:2015dpa,delaCruz:2015raa}.
The number of independent amplitudes is
\bq
 N_{\mathrm{basis}}
 & = &
 \left\{
 \begin{array}{ll}
   \left(n-3\right)!, & n_q \in \{0,1\}, \\
   \left(n-3\right)! \frac{2\left(n_q-1\right)}{n_q!}, & n_q \ge 2. \\
 \end{array}
 \right.
\eq
A possible basis is given for $n_q=0$ by the external orderings
\bq
\label{basis_n_q_0}
 B 
 & = &
 \left\{ 
  \; l_1 l_2 ... l_n \; | \; l_{1}=g_{1}, \; l_{n-1}=g_{n-1}, \; l_n=g_n \;
 \right\}.
 \nonumber
\eq
For $n_q=1$ we may choose
\bq
\label{basis_n_q_1}
 B 
 & = &
 \left\{ 
  \; l_1 l_2 ... l_n \; | \; l_{1}=q_{1}, \; l_{n-1}=g_{n-2}, \; l_n=\bar{q}_1 \;
 \right\}.
 \nonumber
\eq
For $n_q\ge 2$ we may choose
\bq
\label{basis_n_q_2}
 B 
 & = &
 \left\{ 
  \; l_1 l_2 ... l_n \in \mathrm{Dyck}_{n_q} \; | \; l_{1}=q_1, \; l_{n-1} \in \{\bar{q}_2,...,\bar{q}_{n_q}\}, 
 \right. \nonumber \\
 & & \left. l_n=\bar{q}_1 \;
 \right\}.
 \nonumber
\eq
The set $\mathrm{Dyck}_{n_q}$ of Dyck words is defined as follows \cite{Melia:2013bta,Melia:2013epa}:
Let us assume that a quark of flavour $i$ corresponds to an opening bracket ``$(_i$'' and an antiquark of flavour $i$ to a closing bracket ``$)_i$''.
A Dyck word is any word with properly matched brackets,
where closing brackets of type $i$ only match with opening brackets of type $i$.

We denote by ${\mathcal T}(w)$ the set of all ordered tree diagrams with trivalent flavour-conserving vertices and 
external ordering $w$.
This allows two types of vertices: 
A vertex with three un-flavoured particles (``three-gluon vertex'')
and a vertex with one flavoured, one anti-flavoured and one un-flavoured particle (``quark-antiquark-gluon vertex'').
For $n_q=0$ the number of diagrams in the set ${\mathcal T}(w)$ is $(2n-4)!/(n-2)!/(n-1)!$, 
for $n_q>0$ the number of diagrams will depend on $w$. 
Two diagrams with different external orderings are considered to be equivalent, if we can transform one diagram into the other by a sequence of flips.
Under a flip operation one exchanges at a vertex two branches.
We denote by ${\mathcal T}(w_1) \cap {\mathcal T}(w_2)$ the set of diagrams compatible with the external orderings $w_1$ and $w_2$
and by $n_{\mathrm{flip}}(w_1,w_2)$ the number of flips needed to transform any diagram from ${\mathcal T}(w_1) \cap {\mathcal T}(w_2)$ with the external ordering
$w_1$ into a diagram with the external ordering $w_2$.
The number $n_{\mathrm{flip}}(w_1,w_2)$ will be the same for all diagrams from ${\mathcal T}(w_1) \cap {\mathcal T}(w_2)$.
For a diagram $G$ we denote by $E(G)$ the set of the internal edges and by $s_e$ and $m_e$ the Lorentz invariant and the mass corresponding to the internal 
edge $e$.
We denote by ${\mathcal U}$ the set of all unordered tree diagrams with trivalent flavour-conserving vertices.
The number of diagrams in the set ${\mathcal U}$ is $(2n-5)!!/(2n_q-1)!!$.

\section{Colour-kinematics duality}

In this paragraph we review the construction based on colour-kinematics duality \cite{Bern:2010ue,Johansson:2014zca}.
Colour-kinematics duality states that primitive QCD amplitudes can always be brought into a form
\bq
 A_n\left(p,w,\eps\right)
 & = &
 i 
 \sum\limits_{G \in {\mathcal T}(w)}
 \frac{N\left(G\right)}{D\left(G\right)}
 \nonumber \\
 \mbox{with} & &
 D\left(G\right) = \prod\limits_{e \in E(G)} \left(s_e-m_e^2\right),
\eq
where the kinematical numerators $N(G)$ satisfy antisymmetry and Jacobi relations whenever the corresponding colour factors
do.
This will require to decompose diagrams with four-gluon vertices into diagrams with trivalent vertices only.
In general, a solution for the kinematical numerators will not be unique. This freedom is known as 
generalised gauge invariance.
Once the kinematical numerators are constructed, the gravitational amplitude is given by
\bq
\label{def_M_n_v1}
 M_n\left(p, \eps, \tilde{\eps} \right)
 & = &
 \left(-1\right)^{n-3}
 i
 \sum\limits_{G \in {\mathcal U}}
 \frac{N\left(G\right)N\left(G\right)}{D\left(G\right)}.
\eq
Eq.~(\ref{def_M_n_v1}) is independent of generalised gauge transformations.

\section{Generalised KLT-relations}

For $w_1, w_2 \in B$ we define a $N_{\mathrm{basis}} \times N_{\mathrm{basis}}$-dimensional matrix $m$ by
\bq
\label{def_m_n}
\lefteqn{
 m_{w_1 w_2}
 = 
 \left(-1\right)^{n-3+n_{\mathrm{flip}}(w_1,w_2)}
} & & \nonumber \\
 & &
 \sum\limits_{G \in {\mathcal T}(w_1) \cap {\mathcal T}(w_2)} 
 \;\;\;
 \prod\limits_{e \in E(G)} \frac{1}{s_e-m_e^2}.
\eq
We define the momentum kernel $S$ \cite{Kawai:1985xq,BjerrumBohr:2010ta,BjerrumBohr:2010hn,Bjerrum-Bohr:2013bxa,Cachazo:2013gna}
as the inverse of the matrix $m$:
\bq
\label{def_S}
 S & = & m^{-1}.
\eq
Eq.~(\ref{def_m_n}) defines the double-ordered scalar amplitudes $m_n$ 
\bq
 m_n\left(p,w,\tilde{w}\right) & = & m_{w \tilde{w}}.
\eq
The un-ordered gravitational amplitude $M_n$ is given by
\bq
\label{def_M_n}
 M_n\left(p,\eps,\tilde{\eps}\right) & = &
 - i \sum\limits_{w,\tilde{w}\in B}
 A_n\left(p,w,\eps\right) S_{w \tilde{w}} A_n\left(p,\tilde{w},\tilde{\eps}\right).
 \nonumber \\
\eq
For $n_q=0$ eq.~(\ref{def_M_n}) is the well-known 
KLT-relation \cite{Kawai:1985xq} expressed in the basis $B$, relating pure gluon amplitudes in Yang-Mills theory
to graviton amplitudes in gravity.
Eq.~(\ref{def_M_n}) 
in combination with eq.~(\ref{def_m_n}) and eq.~(\ref{def_S})
is the appropriate generalisation for $n_q>0$.
This is a new result.
The essential ingredient is the restriction to diagrams with flavour-conserving vertices in the definition of the set ${\mathcal T}(w)$. 
This implies that for $n=4$ and $n_q=2$ the momentum kernel $S$ differs from the one for $n=4$ and $n_q=0,1$. 
We have checked for all cases with $n \le 8$ that eq.~(\ref{def_M_n}) agrees with eq.~(\ref{def_M_n_v1}).
Starting from $n=8$ and $n_q=4$ the entries of $S$ will involve a non-factorisable polynomial in the denominator. 
The zeros of this polynomial are spurious singularities. 
This can be understood from eq.~(\ref{def_M_n_v1}), since the numerators $N(G)$ can be chosen as local functions. 
To achieve locality for the $n=8$, $n_q=4$ numerators a rearrangement proportional to the
fundamental ``Jacobi'' identity 
\bq
 T^{a_1} T^{a_2} 
 -
 T^{a_2} T^{a_1} 
 -
 i f^{a_1 a_2 b} T^b
 & = & 0
\eq
is necessary for Feynman gauge.

\section{Amplitudes}
\label{sect:amplitudes}

We will write for the gravitational amplitude with the coupling reinserted
\bq
\label{coupling_convention}
 {\mathcal M}_n\left(p,\eps,\tilde{\eps}\right)
 & = &
 \left( \frac{\kappa}{4} \right)^{n-2}
 M_n\left(p,\eps,\tilde{\eps}\right).
\eq
We have $\kappa=\sqrt{32 \pi G_N}$ in the Gau{\ss} unit system.
The perturbative expansion is in $\kappa/4$. This is due to our conventions for the cyclic ordered three-gluon vertrex.
We use
\bq
 i g^{\mu_1\mu_2} \left( p_1^{\mu_3} - p_2^{\mu_3} \right)
 + \mathrm{cyclic \; permutations}.
\eq
In the literature one also finds the convention that the three-gluon vertex has an extra factor $1/\sqrt{2}$, in which case
$\kappa/2$ would  appear in eq.~(\ref{coupling_convention}).

Let us consider the four-point amplitudes involving ``double copies of fermions''. 
The four-point amplitudes are the most relevant ones for phenomenological applications.
We can construct them from the primitive QCD amplitudes. We denote by
\bq
\label{A_4_qggq}
 A_4 & = & 
 A_4\left(q_1^{\lambda_1},g_2^{\lambda_2},g_3^{\lambda_3},\bar{q}_4^{\lambda_4}\right)
\eq
the four-point amplitude for a quark-antiquark pair and two gluons.
The variables $\lambda_i$ label the spin states. 
The quarks may be massive.
The amplitude in eq.~(\ref{A_4_qggq}) is cyclic ordered 
and has poles $1/(2 p_1 p_2)$ and $1/(2 p_2 p_3)$. 
Similar, we denote by
\bq
\label{A_4_qqqq}
 A_4'
 & = &
 A_4\left(q_1^{\lambda_1},q_2'{}^{\lambda_2},\bar{q}_3'{}^{\lambda_3},\bar{q}_4^{\lambda_4}\right)
\eq
the four-point amplitude for two non-identical quark-antiquark pairs.
The quarks may be massive, we denote the mass of $q_1$ and $\bar{q}_4$ by $m$ and the mass of $q_2'$ and $\bar{q}_3'$ by $m'$.
The amplitude in eq.~(\ref{A_4_qqqq}) is cyclic ordered and has 
a pole $1/(2 p_2 p_3 + 2 m'^2)$. 
Let us denote double copies of gluons by $h$ and double copies of quarks/antiquarks by $d$ and $\bar{d}$.
From eq.~(\ref{def_M_n}) we obtain 
\bq
\label{M_4_dhhd}
\lefteqn{
 M_4\left( d_1^{\lambda_1 \tilde{\lambda}_1},h_2^{\lambda_2 \tilde{\lambda}_2},h_3^{\lambda_3 \tilde{\lambda}_3},\bar{d}_4^{\lambda_4 \tilde{\lambda}_4}\right)
 = 
 - i 
 \frac{2 p_1 p_2 \; 2 p_2 p_3}{2 p_1 p_3} 
} & & \nonumber \\
 & &
 A_4\left(q_1^{\lambda_1},g_2^{\lambda_2},g_3^{\lambda_3},\bar{q}_4^{\lambda_4}\right)
 A_4\left(q_1^{\tilde{\lambda}_1},g_2^{\tilde{\lambda}_2},g_3^{\tilde{\lambda}_3},\bar{q}_4^{\tilde{\lambda}_4}\right)
 \nonumber \\
\eq
and
\bq
\label{M_4_dddd}
\lefteqn{
 M_4'\left( d_1^{\lambda_1 \tilde{\lambda}_1},d_2'{}^{\lambda_2 \tilde{\lambda}_2},\bar{d}_3'{}^{\lambda_3 \tilde{\lambda}_3},\bar{d}_4^{\lambda_4 \tilde{\lambda}_4}\right)
 = 
 i 
 \left( 2 p_2 p_3 + 2 m'^2 \right)
} & & \nonumber \\
 & &
 A_4\left(q_1^{\lambda_1},q_2'{}^{\lambda_2},\bar{q}_3'{}^{\lambda_3},\bar{q}_4^{\lambda_4}\right)
 A_4\left(q_1^{\tilde{\lambda}_1},q_2'{}^{\tilde{\lambda}_2},\bar{q}_3'{}^{\tilde{\lambda}_3},\bar{q}_4^{\tilde{\lambda}_4}\right).
 \nonumber \\
\eq
The amplitudes $M_4$ and $M_4'$ are free of double poles. 
In the case of $M_4$, the prefactors $2p_1 p_2$ and
$2p_2p_3$ cancel one of the double poles obtained by squaring $A_4$.
The same is true for $M_4'$.
Here, the prefactor $(2p_2p_3+2m'^2)$ cancels the double pole obtained by squaring 
$A_4'$.
Note that the appropriate momentum kernel for $M_4'$ is the $1 \times 1$-matrix
\bq
 S & = & - \left( 2 p_2 p_3 + 2 m'^2 \right).
\eq
The minus sign is responsible for an attractive $1/r$-potential in the classical limit.
 
\section{Cross sections}
\label{sect:cross_sections}

For the computation of the cross section we have to sum over all spin states of the theory.
In this section we consider for all double copies all four possible spin states $++$, $--$, $+-$ and $-+$.
This has the advantage that the spin sum for ${\mathcal M}$ factorises into two individual spin sums for $A$.
Let us first consider ${\mathcal M}_4$ from eq.~(\ref{M_4_dhhd}).
The spin-summed matrix element squared is given by
\bq
 \left| {\mathcal M}_4 \right|^2
 & = &
 \frac{\kappa^4}{256} 
 \frac{\left(2p_1p_2\right)^2\left(2p_2p_3\right)^2}{\left(2p_1p_3\right)^2}
 \left( \left| A_4 \right|^2 \right)^2,
\eq
where $|A_4|^2$ is the spin-summed matrix element squared obtained from the colour-ordered QCD process
$0 \rightarrow q_1 g_2 g_3 \bar{q}_4$. We have
\bq
\lefteqn{
 \left| A_4 \right|^2
 = } & &
 \\
 & &
 8 \left[
 \frac{2 \left(s-m^2\right)^2}{t^2}
 + \frac{4 s}{t}
 + 3
 + \frac{t+4m^2}{s-m^2}
 + \frac{4 m^4}{\left(s-m^2\right)^2}
 \right].
 \nonumber
\eq
The Mandelstam variables are as usual 
$s=(p_1+p_2)^2$, $t=(p_2+p_3)^2$ and $u=(p_1+p_3)^2$.
For the annihilation cross section $\bar{d}_1 d_4 \rightarrow h_2 h_3$ we obtain
\bq
 \sigma
 & = &
 \frac{4 \pi G_N^2}{4 E_1 E_4 \left| \vec{v}_1 - \vec{v}_4 \right|}
 \left[
   \frac{7}{60} t^2 + \frac{16}{15} m^2 t + \frac{103}{15} m^4 + 8 \frac{m^6}{t}
 \right. \nonumber \\
 & & \left.
   - 4 \frac{m^4}{\chi t} \left(t^2 + 4 m^2 t - 4 m^4 \right)
       \ln\left(\frac{t+\chi}{t-\chi}\right)
 \right],
\eq
with $\chi = \sqrt{t(t-4m^2)}$.
The quantity $4 E_1 E_4 | \vec{v}_1 - \vec{v}_4 |$ denotes the flux factor of the incoming particles $\bar{d}_1$ and $d_4$.
Let us now look at ${\mathcal M}_4'$ from eq.~(\ref{M_4_dddd}).
We have
\bq
\label{Mp4_squared}
 \left| {\mathcal M}_4' \right|^2
 & = &
 \frac{\kappa^4}{256} 
 \left( 2 p_2 p_3 + 2 m'^2 \right)^2
 \left( \left| A_4' \right|^2 \right)^2,
\eq
with
\bq
 \left| A_4' \right|^2
 & = &
 8 \left[
 \frac{2 \left(u-m^2-m'^2\right)^2}{t^2}
 + \frac{2 u}{t} 
 + 1
 \right].
\eq
Let us first look at the process $\bar{d}_1 d_4 \rightarrow d_2' \bar{d}_3'$,
i.e. the annihilation of the pair $\bar{d}_1 d_4$ followed by the creation of a pair $d_2' \bar{d}_3'$ of different flavour.
We obtain for the corresponding cross section
\bq
 \sigma
 & = &
 \frac{4 \pi G_N^2}{4 E_1 E_4 \left| \vec{v}_1 - \vec{v}_4 \right|}
 \sqrt{\frac{t-4m'^2}{t}}
 \left[
   \frac{7}{30} t^2 
 \right. \nonumber \\
 & & \left.
   + \frac{4}{5} \left( m^2 + m'^2 \right) t 
   + \frac{16}{15} \left( m^4 + m'^4 \right)
   + \frac{64}{15} m^2 m'^2
 \right. \nonumber \\
 & & \left.
   + \frac{32}{15} \frac{m^2 m'^2 \left(m^2+m'^2\right)}{t}
   + \frac{32 m^4 m'^4}{5 t^2}
 \right].
\eq
The matrix element of eq.~(\ref{Mp4_squared}) is also relevant 
to the crossing $d_4 d_3' \rightarrow d_1 d_2'$, i.e. the scattering of $d_4$ and $d_3'$.
We obtain for the differential cross section 
with $z = - \cos\theta$
\bq
\label{diff_x_section}
 \frac{d\sigma}{dz}
 & = &
 \frac{4 \pi G_N^2}{4 E_3 E_4 \left| \vec{v}_3 - \vec{v}_4 \right|}
 \sqrt{\frac{P^2}{s}}
 \left[
 4 \frac{\left(s-m^2-m'^2\right)^4}{P^4 \left(z+1\right)^2}
 \right. \nonumber \\
 & & \left.
 - 4 \frac{s\left(s-m^2-m'^2\right)^2}{P^2 \left(z+1\right)}
 + \left(s-m^2-m'^2\right)^2
 \right. \nonumber \\
 & & \left.
 + \frac{1}{16} \left( 4 s - P^2 \left(z+1\right) \right)^2
 \right].
\eq
Here, we set
\bq
 P^2 & = &
 \frac{1}{s}
 \left[ \left(s-m^2-m'^2\right)^2 - 4 m^2 m'^2 \right].
\eq
It is instructive to consider the non-relativistic limit of eq.~(\ref{diff_x_section}).
In the centre-of-mass system one finds 
\bq
\label{non_rel_limit}
 \frac{d\sigma}{dz}
 & = &
 \frac{2 \pi G_N^2 m^2 m'^2}{E^2 \left(z+1\right)^2},
\eq
where $E$ is the total kinetic energy.
This may be compared with the classical Rutherford cross section for the gravitational scattering of
two particles of mass $m$ and $m'$:
\bq
 \left. \frac{d\sigma}{dz} \right.^{\mathrm{Rutherford}}
 & = &
 \frac{2 \pi G_N^2 m^2 m'^2}{4 E^2 \left(z+1\right)^2}.
\eq
The cross section of eq.~(\ref{non_rel_limit}) is a factor $4$ larger than the Rutherford cross section.
The explanation is as follows: We may analyse $A_4'$ and $M_4'$ in the non-relativistic limit. 
In this limit, the internal propagator is almost on-shell
and we may replace the tensor structure by a polarisation sum.
One finds that both polarisations $+$ and $-$ of the internal gluon contribute to $A_4'$ with equal strength.
This implies that $M_4'$ receives equal contributions from all four polarisations $++$, $--$, $+-$ and $-+$.
Therefore $M_4'$ is a factor $2$ larger compared to an amplitude with only $++$ and $--$ exchange.
This gives a factor $4$ in the cross section.

\section{Dark matter}
\label{sect:dark_matter}

Let us recall the main features of dark matter.
Dark matter constitutes roughly $25 \%$ of the energy content of the universe.
Up to now, all evidence for dark matter is gravitational.
The structure formation of the universe indicates that the major part of dark matter is either cold or warm,
thus suggesting massive non-relativistic particles as candidates for dark matter.

The model discussed in this letter might be relevant to the discussion of dark matter.
Double copies of fermions interact only through gravitational 
interactions \footnote{There are other dark matter models, where dark matter particles interact only through gravitational
interactions \cite{Khlopov:1989fj,Hodges:1993yb,Berezhiani:1995am}.}.
The double copies of fermions may be massive and thus non-relativistic.
As they only interact gravitationally, they are dissipationless, compatible with the observation of
dark matter halos around galaxies.

However, as they only interact gravitationally, all cross sections are tiny.
The explanation of the dark matter relic abundance within this model would require a non-thermal production mechanism.
In addition we would like to mention that in the previous section we considered for simplicity 
for all particles all four possible spin states. 
A realistic analysis might require to restrict all states to the equal spin states $++$ and $--$.
Techniques to achieve this with the help of ghosts are discussed in \cite{Johansson:2014zca}.

\section{Conclusions}
\label{sect:conclusions}

In this letter we considered double copies of massless or massive fermions.
These double copies interact only through gravitational interactions.
We constructed the tree-level scattering amplitudes by two different methods:
The first method is based on colour-kinematics duality. Our second method generalises the 
KLT-relations and relates gravitational amplitudes, gauge amplitudes and trivalent scalar amplitudes.
It is remarkable that both methods compute the same quantity.
Double copies of massive fermions can be non-relativistic and might be
relevant to the discussion of dark matter.
We gave explicit expressions for the most relevant cross sections.

\subsection*{Acknowledgements}

S.W. would like to thank Zvi Bern for useful discussions.
L.d.l.C. is grateful for financial support from CONACYT and the DAAD.


\bibliography{/home/stefanw/notes/biblio}
\bibliographystyle{/home/stefanw/latex-style/h-physrev5}

\end{document}